\documentstyle[aps]{revtex}
\begin{document}

\title{Reply to the Comment [1] on``Scaling Laws for a System 
with Long--Range Interactions within Tsallis Statistics'' [2]}

\author{R. Salazar and R. Toral\\
Departament de F\'{\i}sica, Universitat de les Illes Balears\\
Instituto Mediterr\'aneo de Estudios Avanzados (IMEDEA, UIB-CSIC)\\
07071 Palma de Mallorca, Spain.\\
}

\maketitle

\vspace{1cm}

The fact that mean field theory is appropriate to describe an Ising model
with long-range interactions has been already shown by Cannas and 
Tamarit \cite{CT96}.
Although not explicited in our Letter, we have used periodic boundary conditions
in all our simulations, such that the maximum possible distance between two
lattice sites is L/2. 

We make no specific comments about the validity (or lack of validity)
of Boltzmann-Gibbs statistical mechanics to describe long-range systems.
Our paper does nothing but to compare the scaling functions derived
from Boltzman-Gibbs statistics from those derived from the use of Tsallis
entropy.

\end{document}